\documentclass[conference]{IEEEtran}
\usepackage{cite}
\usepackage{amsmath,amssymb,amsfonts}
\usepackage{algorithmic}
\usepackage{graphicx}
\usepackage{textcomp}
\usepackage{xcolor}
\usepackage{siunitx}
\usepackage[nolist]{acronym}
\usepackage{nicefrac}
\usepackage{paralist}
\usepackage{subcaption}
\usepackage{balance}
\usepackage{authblk}
\usepackage{hyperref}
\def\BibTeX{{\rm B\kern-.05em{\sc i\kern-.025em b}\kern-.08em
    T\kern-.1667em\lower.7ex\hbox{E}\kern-.125emX}}
\begin{document}

\title{How Reliable is Smartphone-based \\ Electronic Contact Tracing for COVID-19? \\ {\Large A Look through the Lens of Neighbor Discovery Protocols}}
\renewcommand\Affilfont{\footnotesize}

\author[1]{Philipp H. Kindt}
\author[2]{Trinad Chakraborty}
\author[3]{Samarjit Chakraborty}
\affil[1]{Technical University of Munich, Chair of Real-Time Computer Systems, Munich, Germany. E-Mail: philipp.kindt@tum.de}
\affil[2]{Centre for Medical Microbiology and Virology, Faculty of Medicine, Justus-Liebig University, Giessen, Germany \& \protect\\ German Center of Infection Research, Site Giessen-Marburg-Langen. E-Mail: trinad.chakraborty@mikrobio.med.uni-giessen.de}
\affil[3]{University of North Carolina at Chapel Hill, Department of Computer Science, Chapel Hill, USA. E-Mail: samarjit@cs.unc.edu}

\maketitle

\begin{abstract}
Smartphone-based electronic contact tracing is currently considered an essential tool towards easing lockdowns, curfews, and shelter-in-place orders  issued by most governments around the world  in response to the 2020 novel coronavirus (SARS-CoV-2) crisis. While the focus on developing smartphone-based contact tracing applications or apps has been on privacy concerns stemming from the use of such apps, an important question that has not received sufficient attention is: {\em How reliable will such smartphone-based electronic contact tracing be?} 

This is a technical question related to how two smartphones reliably register their mutual proximity. 
Here, we examine in detail the technical prerequisites required for effective smartphone-based contact tracing. The underlying mechanism that any contact tracing app relies on is called {\em Neighbor Discovery} (ND), which involves smartphones transmitting and scanning for Bluetooth signals to record their mutual presence whenever they are in close proximity. The hardware support and the software protocols used for ND in smartphones, however, were not designed for reliable contact tracing. In this paper, we quantitatively evaluate how reliably can smartphones do contact tracing. We show that irrespective of how well a contact tracing app is designed, because of the limitations stemming from how ND mechanisms are implemented on currently-available phones, they cannot achieve {\em gapless} contact tracing. Our results point towards the design of a wearable solution for contact tracing that can overcome the shortcomings of a smartphone-based solution to provide more reliable and accurate contact tracing. Combining efficacious contact tracing strategies with the available epidemic spread models, will ultimately lead to more accurate estimates on the adoption rates of contact tracing devices that are necessary for containing contagion spread. To the best of our knowledge, this is the first study that quantifies, both, the suitability and also the drawbacks of smartphone-based contact tracing. Further, our results can be used to parameterize a ND protocol to maximize the reliability of any contact tracing app that uses it.

\end{abstract}


\section{Introduction}
\label{sec:introduction}

\noindent 
{\bf Background:}
The global surge of the novel coronavirus SARS-CoV-2 in 2020 has led to a partial, and at some places even a complete lockdown across the world. Since every infected person can potentially cause multiple secondary infections, the solution adopted is to limit social contacts by enforcing social distancing and stay-at-home regimes. This has led to severe economic and other challenges. A promising solution being considered to enable the gradual easing of lockdowns is wireless contact tracing using smartphones. If all relevant previous contacts of a person tested positive for the virus are quickly and reliably identified and isolated, then any further spread of the infection will be reduced.

When modeling the spread of infections in a pandemic, every infected person is considered to infect $R$ others, where $R$ is referred to as the \textit{effective reproduction rate}. Recent studies, e.g.,~\cite{ferretti:20}, suggest that the value of $R$ can be reduced using electronic contact tracing. The extent of this reduction depends on the probability of detecting every individual contact of an infected person. This probability is composed of two factors. The first is the fraction of the population that uses a smartphone-based contact tracing application or app, i.e., its adoption rate. The second factor is determined by the probability with which a smartphone is able to reliably detect a contact. While a lot of current discussion has as focused on the potential adoption rates of various contact tracing apps, it has been implicitly assumed that whenever a smartphone is used, it {\em reliably} registers all relevant contacts. The validity of this reliability assumption has however not been studied so far, and is the focus of this paper. \\


\noindent 
{\bf Neighbor Discovery -- The Contact Tracing Basis:}
Detecting contacts could potentially be done based on location services, such as the global positioning system (GPS). However, besides privacy issues, GPS data is very inaccurate or entirely unavailable inside buildings and therefore of limited use. Hence, devices in vicinity are detected using short-range wireless signals. 
The mechanism that lies at the heart of a smartphone detecting the existence of another smartphone in its vicinity is called \textit{neighbor discovery} (ND). It is based on the phones emitting and scanning for Bluetooth signals, and a successful reception of an emitted Bluetooth signal by another phone and vice-versa will lead to their mutual discovery. 
Any contact tracing app will have to rely on this ND service provided by a smartphone. Because both signal emission as well as reception costs energy, and phones are battery powered, the exact patterns of signal emission and reception are governed by a ND protocol that attempts to balance the discovery latency and the energy expenditure. 
Hence, both, the specific neighbor discovery protocol and the manner in which it is parameterized determine the discovery latency, i.e., how quickly two smartphones will be able to ``discover'' each other when they come in close contact. The protocol and its parameterization also determine its energy consumption, and thereby how long 
a smartphone's battery will last. Other properties, such as the reliability of operation when a large number of phones are in range of reception, are also determined by this.

The original Bluetooth BR/EDR protocol, which existed before \ac{BLE} was released, was primarily designed for  ``pairing'' the phones with other devices like keyboards, computers or Bluetooth speakers for the purpose of data communication. This pairing process is not very time-sensitive, and was not designed for reliable and sustained contact tracing, as currently envisioned against COVID-19. In any traditional pairing process, if the pairing is not successful then the user resets one or both the devices and attempts again, and repeats the process until the two devices are successfully paired for communication. However, this manual intervention and checking whether a pairing has been successful is not feasible in the case of contact tracing, where two or more phones are {\em always} expected to reliably ``pair'' on their own whenever they are in close proximity even for relatively short amounts of time. In contrast, the \ac{BLE} protocol has been explicitly designed for continuously scanning in the background and is therefore the primary choice for \ac{ND} on smartphones. However, not all capabilities and degrees of freedom \ac{BLE} offers for \ac{ND} are available on a phone. In addition, apps commonly use \ac{BLE} in the fashion of Bluetooth BR/EDR, where scanning for other devices is triggered manually.\\

\noindent 
{\bf Limitations of Smartphones:}
Hence, when a smartphone is used for contact tracing, a relevant question is: whenever two or more people come in contact with each other, what is the probability of their respective smartphones being able to record such a contact? If the duration of the contact is very brief (e.g., a couple of seconds), would such contacts still be detected? What is the minimum duration of a contact that will be reliably detected? Will smartphones be able to detect the \textit{type} and proximity of the contact? For example, were the two subjects within 1.5 meters of each other or at less than 0.1 meters? Did they touch each other? The answers to these questions in the context of smartphone-based contact tracing is not clear. Nevertheless, both among medical professionals, and even software developers (who might not have the relevant background in communications technology and computer hardware to appreciate the neighbor discovery process in smartphones), contact tracing apps are now being seen as the holy grail for this problem. However, what a contact tracing app might or might not be able to do will be fundamentally limited by the underlying hardware and protocol configuration of ND mechanisms supported by the smartphone device. 

Besides the success probability of the ND procedure itself, the distance estimation is highly susceptible to errors. This estimation is based on sensing the attenuation of the wireless signal, which in free space correlates with the distance between two devices. However, human tissue attenuates these signals much more, which can make distance estimations to be erroneous. Therefore, the accuracy of distance estimation will depend on the particular positioning of the devices detecting each other, e.g., are they in one's hands, in the pocket, or in a bag? The accuracy will also depend on the relative orientation of the people carrying the phones. 
In addition, there are different other sources of error, e.g., variations among devices, signal reflexions in the environment and frequency-dependent receiver sensitivities. As a result, contacts that are not relevant could be misclassified as relevant, (viz., result in \textit{false positives}), and actually relevant contacts might not be registered. Since both testing (for SARS-CoV-2) and isolation are expensive, there is a strong need to avoid too many false positives, which would lead to testing and/or isolating a large number of uninfected people. Hence, a sufficient ``safety margin'' needs to be built into the distance estimation procedure, which in turn will lead to a higher rate of unregistered relevant contacts. \\

\noindent
{\bf Contributions of This Paper:} 
In this paper, we attempt to address the above questions. We systematically evaluate the suitability of ND configurations in commercially-available smartphones for the purpose of electronic contact tracing. Our study exposes the fundamental limits that any smartphone will have, no matter which contact tracing app is used. 
We show that distance estimation can only be done with limited accuracy. 
We also discuss issues related to how a smartphone should be used for effective contact tracing. From this we conclude that for many users -- especially susceptible population groups, such as the elderly, who may not be comfortable with technology -- a smartphone-based solution will be of limited value. 

We finally argue that for ``gapless'' contact tracing, smartphones are not suitable. 
Results from disease spread models (e.g., \cite{ferretti:20}) show that contact tracing using smartphones can considerably reduce the reproduction rate $R$, even with partial adoption of contact tracing apps. Our study -- on the reliability of the ND process in smartphones -- when combined with such disease spread models would result in a more accurate estimation of the necessary adoption rates in order to achieve the desired containment. It can also quantify how improving the reliability of tracing mechanisms (cf. smartphone-based ones) would lead to a better decline of a spread. 

For example, the models in \cite{ferretti:20} indicate that even when the delay between patients developing symptoms until their identification and isolation is only one day, the success probability for contact tracing needs to exceed at least $\SI{60}{\percent}$ (for the lowest estimate of the basic reproduction number $R_0$ in~\cite{ferretti:20}) in order to contain a spread. Given that contacts cannot be recorded with $\SI{100}{\percent}$ reliability using smartphones (as we show in this paper), the adoption of contact tracing apps has to be considerably larger than $\SI{60}{\percent}$ to contain the spread (i.e., to ensure that $R < 1$). Given that high adoption rates appear to be unrealistic (only a very small fraction of the population has adopted such apps, including in countries like Singapore, where the adoption rates were expected to be much higher), smartphone-based contact tracing would have to be augmented by additional social distancing measures, which are associated with their own challenges. Alternatively, we need more reliable tracing mechanisms compared to what currently-available smartphones would be able to facilitate. 

Driven by this insight that {\em detection reliability} is a critical variable, we propose a wearable solution for contact tracing, such as an electronic bracelet. We discuss its design and advantages and argue that 
a wearable solution can potentially mitigate all the major limitations  that smartphones suffer from. While such dedicated wearable solutions are yet to be developed and tested in practice, they seem to be a promising alternative to smartphone-based contact tracing. \\

In addition, from our performance evaluation also follows which parameterizations offered by Android should be used for contact tracing apps. To the best of our knowledge, no prior data on which of the available setting performs best for this purpose has been known. Therefore, our results can be directly used for improving the performance of contact tracing apps.\\

\noindent{\bf Summary and Organization:}
In summary, our main contributions are as follows. (a)~We debunk the potential impression that smartphones can reliably conduct contact tracing and the only obstacle is their adoption (stemming from issues such as privacy and security). (b)~Towards this, we lay the foundations for quantifying the reliability and accuracy of contact tracing when using currently-available smartphones. These, when combined with available epidemic spread models, can lead to a better estimation of adoption rates necessary for containment, and the required speed with which potentially infected people should be tested and isolated. (c)~We show why the limited reliability of any smartphone-based contact tracing app, and its other shortcomings, can be addressed using an alternative wearable solution, such as an electronic bracelet. 

The rest of this paper is organized as follows. In the next section, we briefly describe the theoretical foundations of energy- and latency-optimal solutions for contact tracing. In Section~\ref{sec:smartphones}, the performance of contact tracing using currently-available {\tt iOS} and {\tt Android} smartphones is evaluated. In Section~\ref{sec:wearable}, we propose a wearable solution for gapless contact tracing, which could overcome the shortcomings of smartphone-based approaches.
In Section~\ref{sec:concluding_remarks} we conclude that smartphone-based tracing solutions, while technically feasible, do not allow reliable detection and classification of all contacts.
\section{Neighbor Discovery}
\label{sec:neighbor_discovery}
As discussed in the previous section, the mechanism underlying wireless contact tracing is called {\em neighbor discovery} (\textit{\ac{ND}}).
In this section, we briefly describe the \ac{ND} procedure and its performance in general. The goal of this section is to illustrate the design space of \ac{ND} protocols, which involves multiple trade-offs, e.g., latency versus energy consumption versus resilience in crowded situations. Any smartphone application for contact tracing will build on a restricted version of this procedure. We study such restrictions and evaluate the resulting performance in Section~\ref{sec:smartphones}.

Let us first consider two wireless devices that are unaware of their mutual presence, but would like to ``discover'' each other as soon as they are in close proximity. One of them acts as a {\em sender} and the other as a {\em receiver}. The sender continually broadcasts beacons, while the receiver continually listens to the channel for certain time intervals.
All transmissions and receptions are scheduled following a certain pattern that repeats itself after a certain time. The receiver has discovered the sender, once a beacon is sent within a reception window of the receiver. 

The main reason behind both transmission and listening being \textit{continual} and not \textit{continuous} is energy. There is an energy cost for both sending a beacon and for listening to the channel. Another reason, which might hold for some devices, is the parallel execution of other tasks. For example, the receiver might communicate with a third device when not listening for incoming beacons. When not sending or receiving, the devices go to a \textit{sleep} or power-down mode in order to save energy. Since smartphones are battery-powered, saving energy to ensure that the phone does not have to be charged too often is of crucial importance and greatly affects its usability.  From the perspective of \textit{\ac{ND}}, the energy consumption of each device is determined by the fraction of time spent on transmission or reception, i.e., by the \textit{duty-cycle} for transmission $\beta$ and for reception $\gamma$. For a \textit{\ac{ND}} protocol to be \textit{efficient} the goal is to identify a transmission and reception pattern that, for a given tuple $(\beta, \gamma)$, minimizes the time until a beacon is guaranteed to overlap with a reception window in the worst case. In other words, given transmission and reception energy budgets, which transmission and reception patterns will minimize the \textit{worst-case discovery latency}? 

Clearly, in the context of contact tracing, the neighbor discovery procedure should guarantee discovery within a short time interval, i.e., it should be able to register contacts even when two people come in close proximity for relatively short intervals of time, e.g., when shopping at a grocery store, or jogging at a park. However, being able to do so should not quickly drain the battery of the device. While \ac{ND} protocols have been routinely used in a variety of devices, ranging from phones, to laptops, speakers and keyboards, it is only very recently that we have a good understanding of how to exactly determine optimal beacon transmission and reception patterns. In particular, what is the lowest discovery latency $L$ that any receiver with a duty-cycle of $\gamma$ and any sender with a duty-cycle of $\beta$ can achieve was determined in~\cite{kindt:19} and is as follows. 
\begin{equation}
\label{eq:optPerformance}
L = \left \lceil\frac{1}{\gamma}\right\rceil \cdot \frac{\omega}{\beta} + \omega
\end{equation}
where $\omega$ is the transmission duration of a beacon. 
For example, if we require that only contacts that last for at least $5$ seconds are relevant for disease transmission and need to be registered, then $L = \SI{5}{s}$. In order to realize this with a beacon length of $\omega = \SI{40}{\micro s}$, both devices need to be active for about $\beta = \gamma = \SI{0.28}{\percent}$ of their time.

Using which transmission and reception patterns can this performance be achieved? The only known patterns of beacon transmission and reception that achieve optimal discovery latencies are based on {\em periodic intervals}~\cite{kindt:20b, kindt:17a}. They work as follows.
One device periodically broadcasts beacons with a period $T_a$, whereas the other device switches on its receiver for a window of $d_s$ time-units after every $T_s$ time-units. This scheme -- with some minor modifications that we describe below -- is used by \ac{BLE} protocols implemented inside smartphones.

Recall that for one sender and one receiver, the optimal discovery latency for a given energy-budget is known (cf. Equation~\ref{eq:optPerformance}), and this latency guaranteed in $\SI{100}{\percent}$ of all discovery attempts. But the moment both devices act as both senders and receivers, the probability of discovery within the same time interval $L$ now drops. With additional devices being within the range of reception, this probability drops even further. Why is this so, is explained below. 

Consider two devices \textit{A} and \textit{B},  each of which both receives and also transmits. Since they will both run the same firmware or software, both will have the same values of the tuple $(T_a, T_s, d_s)$. Since $T_a$, $T_s$ and $d_s$ are designed such that a beacon of one device should overlap with a reception window of another device, and since the tuple $(T_a, T_s, d_s)$ is identical on both the devices, a beacon of e.g., device \textit{A} will also overlap with a reception window of \textit{A} once every worst-case latency $L$. However, the radio cannot receive and transmit simultaneously at the same time. Hence, the affected reception window needs to be interrupted by one beacon transmission duration $\omega$ plus the time needed by the radio to switch between reception and transmission and vice-versa. Now if \textit{B} transmits a beacon that overlaps with this affected reception window of \textit{A}, it might not be received, since \textit{A} is transmitting or switching between reception and transmission and therefore not listening to the channel. 

Besides the beacon transmission rate, which we describe below, the probability of this failure also depends on the length and hence the transmission duration of a beacon. The more bytes per beacon are transmitted, the higher will be the probability of failed discoveries.
The minimal data to be exchanged by two contact tracing devices in range is a device identifier. To be able to provide at least one unique ID for each device, $\SI{4}{byte}$ or more are required. In addition, for detecting an incoming beacon, a preamble of at least $\SI{1}{byte}$ needs to be added for technical reasons on most radios.  
We hence assume that a $\SI{5}{byte}$-packet needs to be transmitted for contact tracing. For a pair of devices, existing near-optimal approaches~\cite{kindt:20b} can guarantee a worst-case latency of $\SI{5}{s}$ in around $\SI{99.98}{\percent}$ of cases (in contrast to $\SI{100}{\percent}$ when there is only one sender and one receiver), with an energy consumption close to the theoretical optimum. However, as soon as the number of devices increases, this success probability will be reduced even further, as we describe next.

\subsubsection{ND in Crowded Scenarios}
While the success rate is $\SI{99.98}{\percent}$ when only two devices are in close proximity, this rate rapidly drops for a larger number of devices (cf.~\cite{kindt:20b} for a study on this). The mechanism that causes this drop is multiple devices sending beacons at the same point in time. Such overlapping beacons collide and fail to be received. The fact that any two consecutive beacons are spaced by $T_a$ time-units intensifies this problem, since if one pair of beacons from two devices collide, all other pairs of the same device pair will also collide.

There are multiple situations that are of potential relevance for virus transmission, in which a larger number of people are in vicinity of each other. For example, consider a crowed public bus or even a ski gondola. As a worst-case scenario, the maximum density of crowds  without squashing and tilting the human body has been estimated to be $\SI{6}{persons/m^2}$~\cite{oberhagemann:12}. The German government considers contacts within a transmission distance of $\SI{1.5}{m}$ as relevant. If we assume that a radio will have a range of $\SI{2}{m}$ for safely covering the required distance of $\SI{1.5}{m}$, the worst-case number of people/phones in a collision domain is 75. Hence, using known approaches for realizing $L$ using Equation~\ref{eq:optPerformance}, e.g., ~\cite{kindt:20b}, about $\SI{35}{\percent}$ of all discovery attempts will fail. This implies that a significant number of contacts will not be registered. Clearly, countermeasures need to be taken, which trade energy consumption or discovery latency against success probability. In other words, if some increase in the worst-case latency $L$ (again, which is only reached in \SI{100}{\percent} of all cases for only one sender and one receiver) is tolerated, then the success probability when multiple devices need to discover each other can be increased.
In particular, the following two techniques can be used for making neighbor discovery protocols more robust against collisions. Both of them are currently used in smartphones. \\

\par\noindent\textbf{Reducing the Channel Utilization: } The probability of collisions is given by the fraction of time each device utilizes the channel, i.e., $\beta$. Hence, if fewer beacons are sent at any point in time, then the collision probability decreases. As a drawback, since beacons are then sent less frequently, the worst-case latency $L$ will increase, or the duty-cycle for reception $\gamma$ needs to be increased for compensating for the reduced $\beta$ (see Equation~\ref{eq:optPerformance}).
Reducing $\beta$ and in turn increasing $\gamma$ for reaching the same $L$ will, however, increase the overall energy consumption. In practice, when $T_a$ is increased to reduce collisions, $d_s$ needs to be increased for reaching the same worst-case latency, such that the overall energy consumption of every device is increased.
	
For example, when choosing a reduced channel-utilization of $\beta^\prime = \nicefrac{1}{4} \cdot \beta$ and an increased duty-cycle for reception of $\gamma^\prime = 4\cdot \gamma$, the failure probability for the protocol described above when simultaneously discovering $75$ devices can be reduced from $\SI{35}{\percent}$ to about $\SI{10}{\percent}$ without increasing $L$. On a radio in which transmission consumes the same current as reception, the lowest worst-case latency for a given energy budget $\eta = \beta + \gamma$ is achieved for $\beta = \gamma = \nicefrac{1}{2}(\beta + \gamma)$~\cite{kindt:19}. Reducing the channel utilization to $\beta^\prime$ while increasing $\gamma$ to $\gamma^\prime$ will hence lead to an increased sum $\eta^\prime = \beta^\prime + \gamma^\prime = \nicefrac{17}{8}\cdot \eta$ and therefore to an increased energy consumption, while leading to the same $L$. As we will show in Section~\ref{sec:smartphones}, smartphones use a very low channel utilization $\beta$, thereby achieving low collision rates at the expense of increased energy consumption. \\

\par\noindent\textbf{Decorrelation: } A technique used by the \ac{BLE} protocol, which is used in smartphones, is decorrelation. It will further increase the worst-case latency $L$, but reduces the chance of multiple subsequent colliding beacons. 

Instead of sending beacons with periodic intervals, two consecutive beacons can be separated by a fixed amount of time plus a certain random component. For nevertheless guaranteeing the same worst-case latency $L$, the reception duty-cycle $\gamma$ needs to be increased. Consider for example a configuration where $T_a = d_s - \omega$, which has been shown to be a configuration that reaches the smallest possible worst-case latency~\cite{kindt:19}. Here, a beacon will coincide with \emph{every} scan window, since the distance between two consecutive beacons does not exceed the length of a reception window. If now $T_a$ is additionally increased by a random component $\rho \in [0,b]$, $d_s$ needs to be extended by $b$ time-units for ensuring that a beacon will still fall into every scan window. Otherwise, another beacon has to overlap with a later scan window for realizing discovery, thereby increasing the worst-case latency $L$. As a result, the collision rate will only slightly reduce (since the effective distance between two successive beacons is increased to $T_a$ plus the mean value of $b$, which leads to a reduced channel utilization $\beta$), but the collision probabilities of the individual beacons become decorrelated from each other. In other words, if one pair of beacons from two different devices collide, a later pair of beacons only collides with a probability below 1. If now multiple beacons overlap with one or more scan windows of a remote device, there is an increased chance that one of them will not collide and hence, the success probability increases. For strategies based on decorrelation and multiple overlapping beacons, configurations that optimize the trade-off between failure probability, discovery latency and energy have not yet been sufficiently studied in the literature and remain unknown.

In summary, there are multiple degrees of freedom for optimizing the \ac{ND} procedure, and identifying the optimal trade-off between discovery latency, energy consumption and success probability is crucial for efficient and reliable contact tracing. Unfortunately, on a smartphone, these degrees of freedom are not exposed to a contact tracing app and several protocol parameters are already predetermined. In light of this, we examine the performance that is achievable using smartphones in the next section.
\section{Neighbor Discovery on Smartphones}
\label{sec:smartphones}

 In the previous section, we have described the degrees of freedom when designing \ac{ND} protocols in general. The \acf{BLE} protocol used for contact tracing on smartphones restricts these degrees of freedom, and the Android and iOS operating systems further impose restrictions on the \ac{BLE} operation. In this section, we first describe these restrictions that limit the design space when \ac{BLE} is used, and then elaborate on the additional restrictions imposed by Android and iOS. We then study the performance achieved under these restrictions and evaluate whether it is sufficient for reliable contact tracing. Our goals are i)~assessing the tracing performance on existing smartphones, ii)~identifying the most suitable parameterizations on smartphones, and iii)~showing  how a dedicated device (e.g., a wearable) could exploit the availability of the entire design space for improving contact tracing performance.
 
 \subsection{Restrictions in Smartphones}
 \label{sec:constraints}
 \subsubsection{Bluetooth Low Energy}
 \label{sec:ble}
In \ac{BLE}, so-called \textit{advertising events}, within which beacons are sent, are scheduled with a period of $T_a$~\cite{bleSpec52}. Reception windows of length $d_s$ are repeated with a period of $T_s$. Hence, it shares similarities with the neighbor discovery protocol described above. However, $T_a$ is composed of a static part $T_{a,0}$ plus a random delay of $\rho \in [0, \SI{10}{ms}]$. 
Each advertising event consists of a sequence of three beacons. Each of them is sent on a dedicated wireless channel, and these three channels are the same for all devices and advertising events. Optionally, some of these beacons can be omitted in every event, such that e.g., only one channel is used. In this paper we assume that all three channels are used, since this miminizes the discovery latency and is also the only option supported by smartphones. The receiver toggles between the three different channels after each instance of $T_s$. While the values of $T_{a,0}$, $T_s$ and $d_s$ can be chosen freely within a large, quasi-continuous range, the random delay cannot be optimized, since its maximum value is fixed. The 3-channel procedure increases the duty-cycle for transmission $\beta$ by a factor of 3 (since 3 beacons are sent every $T_a$ time-units), and this increased duty-cycle leads to essentially the same discovery latency as the original one. However, it makes the discovery more robust compared to when a subset of these three channels is used.
Finally, the minimal beacon length when broadcasting \ac{BLE} beacons is $\SI{16}{byte}$, because the \ac{BLE} protocol requires multiple packet overheads. In particular, $\SI{3}{byte}$ are used for a \ac{CRC} and $\SI{4}{byte}$ are used as a device identifier. The remaining overhead bytes are of no additional use for contact tracing. This further increases the energy consumption for transmission by a factor of 2-3$\times$ compared to the protocol we described in the previous section.
 
In summary, compared to an energy-optimal protocol, \ac{BLE} introduces an overhead of a factor of approximately $6\times$ to the energy spent for transmission. An additional overhead to $\gamma$ is caused by the random delay $\rho$, since $d_s$ needs to be increased for guaranteeing the same worst-case latency.
The range of the random delay is fixed. However, no or a very small range of possible random delays would probably perform best for a pair of devices, where collisions play a negligible role, while larger delays than specified in \ac{BLE} might further improve the reliability in crowded scenarios where multiple phones are present and send beacons.
On the other hand, some of these overheads improve the reliability of the discovery procedure. In particular, the random delay increases the success probability in crowded scenarios. The \ac{CRC} allows detecting beacons that collided because of concurrent transmissions from multiple devices, which improves the accuracy of distance estimation (which we describe later). The operation on three channels might allow for a later successful discovery when there is interference from non-BLE devices on one of these channels.

 \subsubsection{Android and iOS}
 \label{sec:android}
 \begin{table}[t]
 \small
 \begin{tabular}{l|c|c|c}
 		\textbf{Mode} &$\mathbf{T_{a,0}}$& $\mathbf{d_s}$ & $\mathbf{T_s}$ \\
 		\hline
 		ADVERTISE\_MODE\_LOW\_LATENCY & $100$ & - & - \\
 		ADVERTISE\_MODE\_BALANCED & $250$  &- & -\\
 		ADVERTISE\_MODE\_LOW\_POWER & $1000$ & - & - \\
 		SCAN\_MODE\_LOW\_POWER & - & $512$&$5120$ \\
 		SCAN\_MODE\_BALANCED & - & $1024$&$4096$ \\
 		SCAN\_MODE\_LOW\_LATENCY & - & $4096$&$4096$ \\
 	\end{tabular}

 \caption{Parameter settings supported by Android in [ms].}
 \label{tab:sttings_android}
\end{table}
\begin{figure}[b]
	\centering
	\includegraphics[width=\linewidth]{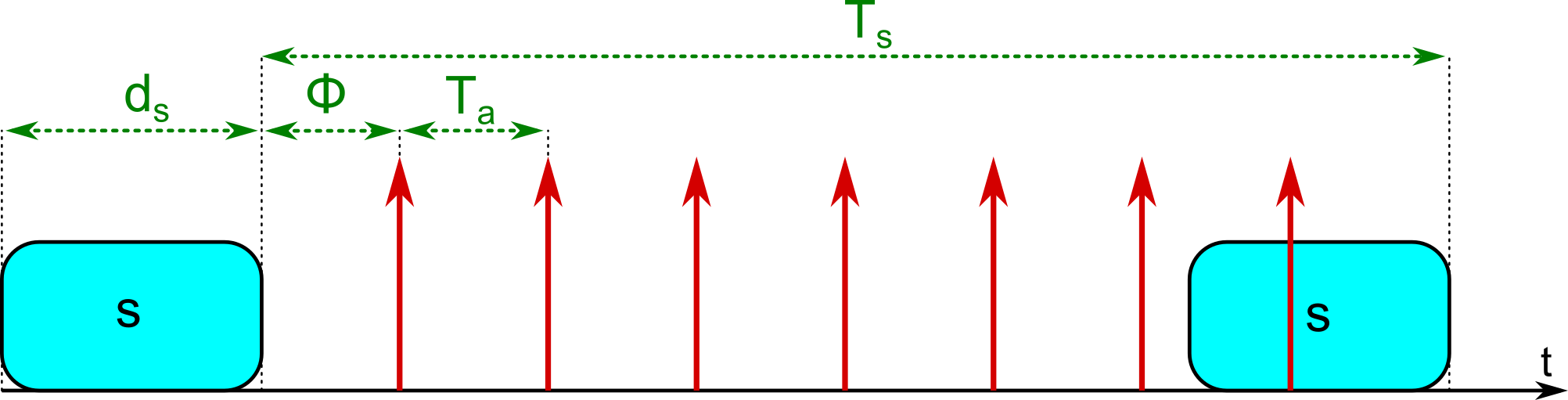}
	\caption{Sequence of beacons (red arrows) with $T_a \leq d_s$ that falls with an offset of $\Phi$ into the scan interval $T_s$, which separates two scan windows (\textit{S}) from each other.}
	\label{fig:Taltds} 
\end{figure}
We have already discussed how \ac{BLE} constraints the designs space of the \ac{ND} procedure. While \ac{BLE} allows essentially any configuration for the tuple $(T_{a,0}, T_s, d_s)$, Android constrains the design space to 3 different settings that affect $(T_s,d_s)$. In addition, Android provides a \textit{batch mode}, where multiple discovered devices are reported jointly with a certain delay. This mode might allow for 3 additional configurations of $(T_s, d_s)$. We do not consider these configurations in this paper, since we could not find a complete documentation covering this feature. In addition to the settings for scanning, Android provides another 3 different settings that affect $T_{a,0}$. 

On a smartphone, fixed values for these parameters are not feasible, because the smartphone hardware might be forced to vary them during runtime. First, the Bluetooth \ac{SoC} might need to carry out advertising and scanning in parallel to other tasks, e.g., streaming audio to a wireless headphone. As a result, the points in time when both tasks need to be served might overlap. Since the device can carry out only one task at a time, some scheduling is needed to resolve this conflict, which might require adapting the values of $T_a, T_s$ and $d_s$ online. In other words, the effective values of $T_a, T_s$ and $d_s$ that are used would be different from the ones that were provided by the operating system's API and were chosen by a contact tracing application. In addition, many smartphones share certain hardware components, e.g., the radio and/or the antenna, for realizing different wireless protocols. The \ac{SoC} or antenna that is responsible for BLE might also be used for WiFi-communication. For example, the Samsung Galaxy S10 smartphone uses the same \ac{SoC} for IEEE 802.11 (WiFi) and Bluetooth~\cite{techinsights:20}. However, it is not possible to e.g., transmit a WiFi frame and a \ac{BLE} beacon at the same time. Indeed, \acp{SoC} that combine \ac{BLE} and WiFi use arbitration mechanisms for this purpose~\cite{silabs:20}. 
The exact method using which such conflicts are resolved in a certain smartphone model is, to the best of our knowledge, not known, since no generic documentation is available on this. It depends on the particular \ac{SoC} used and hence potentially varies between different smartphone models. In general, there are two possibilities for resolving such scheduling conflicts. First, the parameters $T_{a,0}$, $T_s$ and $d_s$ could be chosen such that no advertising packet or scan window overlaps with any other task. Second, if this is not possible, an advertising packet could be skipped, sent earlier/later, or the parameters $T_{a,0}$, $T_s$ and $d_s$ might be altered repeatedly on a short-term basis. 

The values of $T_a, T_s$ and $d_s$ used in Android smartphones are not specified in the official documentation -- the reason for this might lie in the potential need for short-term changes described above. 
However, Android is an open source software, and information that is not provided in its specification documents can be looked up directly from the source code. The Android source code contains different tuples of values $(T_{a,0}, T_s, d_s)$, which are summarized in Table~\ref{tab:sttings_android}.
A pair of values for $T_s$ and $d_s$ can be selected by an application by choosing one of the SCAN\_MODE settings, whereas the value of $T_{a,0}$ can be selected by using one of the ADVERTISE\_MODE settings. 

We could experimentally observe that the $T_{a,0}$-values given in  Table~\ref{tab:sttings_android} are used in smartphones in practice.
However, especially in the presence of other Bluetooth tasks running in parallel (e.g., streaming audio to a wireless headphone), 
the actual values might change in an unpredictable manner. The same holds true for $T_s$ and $d_s$. In addition, different smartphone manufactures might use other values than those from Table~\ref{tab:sttings_android} in their adapted Android versions. Furthermore, they might change in future versions of Android. Indeed, different values for Android are known from previously published work~\cite{siva:19}, indicating that they have been changed in the past.

For iOS, no parameterizations that are being used are documented. Since iOS is a closed source software, we also cannot obtain them from the source code. However, Apple recommends certain values of $T_{a,0}$ for gadgets communicating with iOS devices in~\cite{apple:19}. We therefore assume that iOS devices themselves will also use them, as long as there are no scheduling conflicts. In particular, the following advertising intervals $T_{a,0}$ are recommended for iOS, while no data are given for $T_s$ and $d_s$: $\SI{152.5}{ms}$, $\SI{211.25}{ms}$, $\SI{318.75}{ms}$, $\SI{417.5}{ms}$, $\SI{546.25}{ms}$, $\SI{760}{ms}$, $\SI{852.5}{ms}$, $\SI{1022.5}{ms}$,  $\SI{1285.0}{ms}$. 

In summary, while there are no guaranteed parameter values used by \ac{ND} protocols on smartphones, we can assume that the ones we have described above are used in \textit{most cases}, e.g., when no other Bluetooth communication takes place in parallel.  
In the next section, we evaluate the energy consumption, discovery latency, and success probabilities that can be expected from these values. We also discuss whether the resulting performance is acceptable for contact tracing.

\subsection{Performance Evaluation}
In this section, we evaluate the performance that can be achieved under the constraints of \ac{BLE} on Android. We thereby assume that the values from Table~\ref{tab:sttings_android} for Android and the values of $T_{a,0}$ we have described for iOS are used. As we have explained above, these values might change due to various reasons. We will therefore also discuss how such variations might impact the performance.
Since there are multiple other uncertainties, e.g., different BLE \acp{SoC} used on different smartphones, and the lack of  specifications, 
we thereby estimate the expected bandwidth of performance whenever appropriate. 

In all our evaluations, we assume a packet length of $\SI{16}{byte}$, which is the minimal length supported by the \ac{BLE} standard. This already comprises the length of a MAC address that could be used as a device ID for contact tracing. If additional data are to be sent on top (e.g., a random ID in addition to the MAC address), the packet length is increased, which impacts the energy consumption and collision probability, but not the discovery latency.

\subsubsection{Discovery Latency}
We now study the discovery latency achieved a)~by two Android phones discovering each other, and b)~by an Android phone discovering an iOS phone. 
The situation in which an iOS device discovers an Android phone cannot be evaluated, since no data on $T_s$ and $d_s$ used by iOS are available.

Let us have a closer look on the values from Table~\ref{tab:sttings_android} and the $T_{a,0}$-values we have outlined for iOS.
Most of them fulfill $T_{a} < d_s$. This situation is illustrated by Figure~\ref{fig:Taltds}. After two devices come within their reception range, the first advertising beacon of one device coincides with an arbitrary instance of the scan interval $T_s$ of the other device with an offset of $\Phi$ time-units. $\Phi$ is a random value between $0$ and $T_s$. Since $T_a < d_s$, the latency measured from the first beacon to the successfully received one is limited by roughly $T_s - d_s$ time-units. Because both devices might have been brought into range by up to $T_a + \SI{10}{ms}$ before the first beacon was sent, the worst-case latency is bounded by approximately $T_s - d_s + T_{a,0} + \SI{10}{ms}$ (cf. Figure~\ref{fig:Taltds}). As an example, for the SCAN\_MODE\_LOW\_POWER setting and a value of $T_{a,0} =$ $\SI{100}{ms}$, the worst-case latency is approximately $\SI{4718}{ms}$.

On the other hand, for some other settings, e.g., ADVERTISE\_MODE\_LOW\_POWER in combination with SCAN\_MODE\_LOW\_POWER, $T_a > d_s$. Here, the first scan window might be missed in some cases, and only a later scan window can lead to a successful discovery. For such parameterizations, the latencies might be finite or infinite, depending on $T_{a,0}$, $T_s$ and $d_s$ (see \cite{kindt:15} for details). In any case, since $T_s$ always exceeds $\SI{4}{s}$ in all scanning settings on Android, and since discovery can only be guaranteed after multiple instances of $T_s$, the resulting latencies are significantly larger than for parameterizations with $T_a < d_s$. In contrast, in the SCAN\_MODE\_LOW\_LATENCY setting, the device scans continuously, with only short interruptions for changing the channel. Hence, discovery is successful within $T_{a,0} + \SI{10}{ms}$ time-units, since every beacon is received almost instantaneously. For example, for the ADVERTISE\_MODE\_LOW\_LATENCY setting, the worst-case latency would be $\SI{110}{ms}$. 
 
\begin{figure*}[tb]

\centering
	\begin{subfigure}[t]{0.99\columnwidth}
		\centering
		\includegraphics[width=\linewidth]{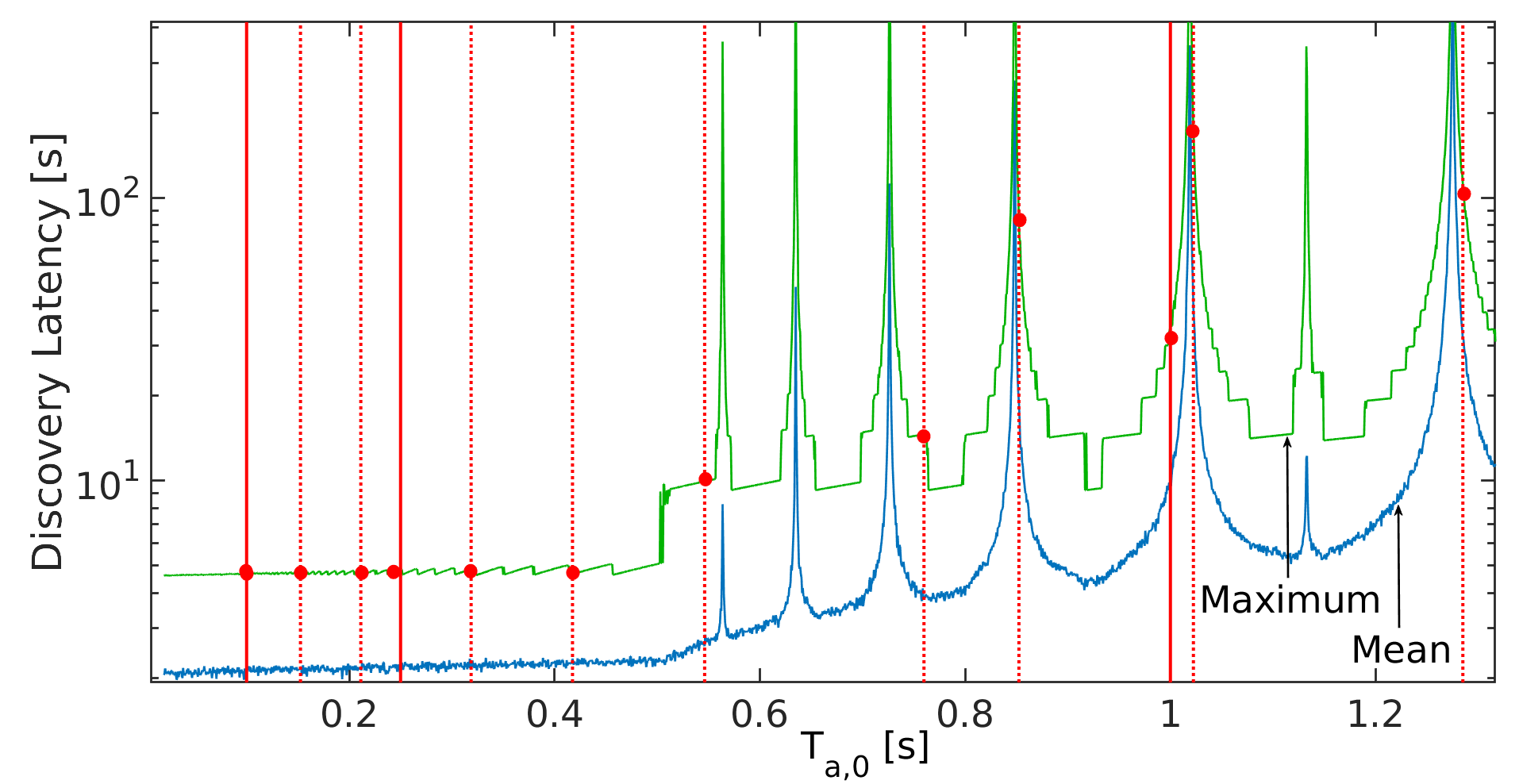}
		\caption{SCAN\_MODE\_LOW\_POWER}
		\label{fig:latenciesAndroidLowPower}	
	\end{subfigure}
	\quad
	\begin{subfigure}[t]{0.99\columnwidth}
		\centering
		\includegraphics[width=\linewidth]{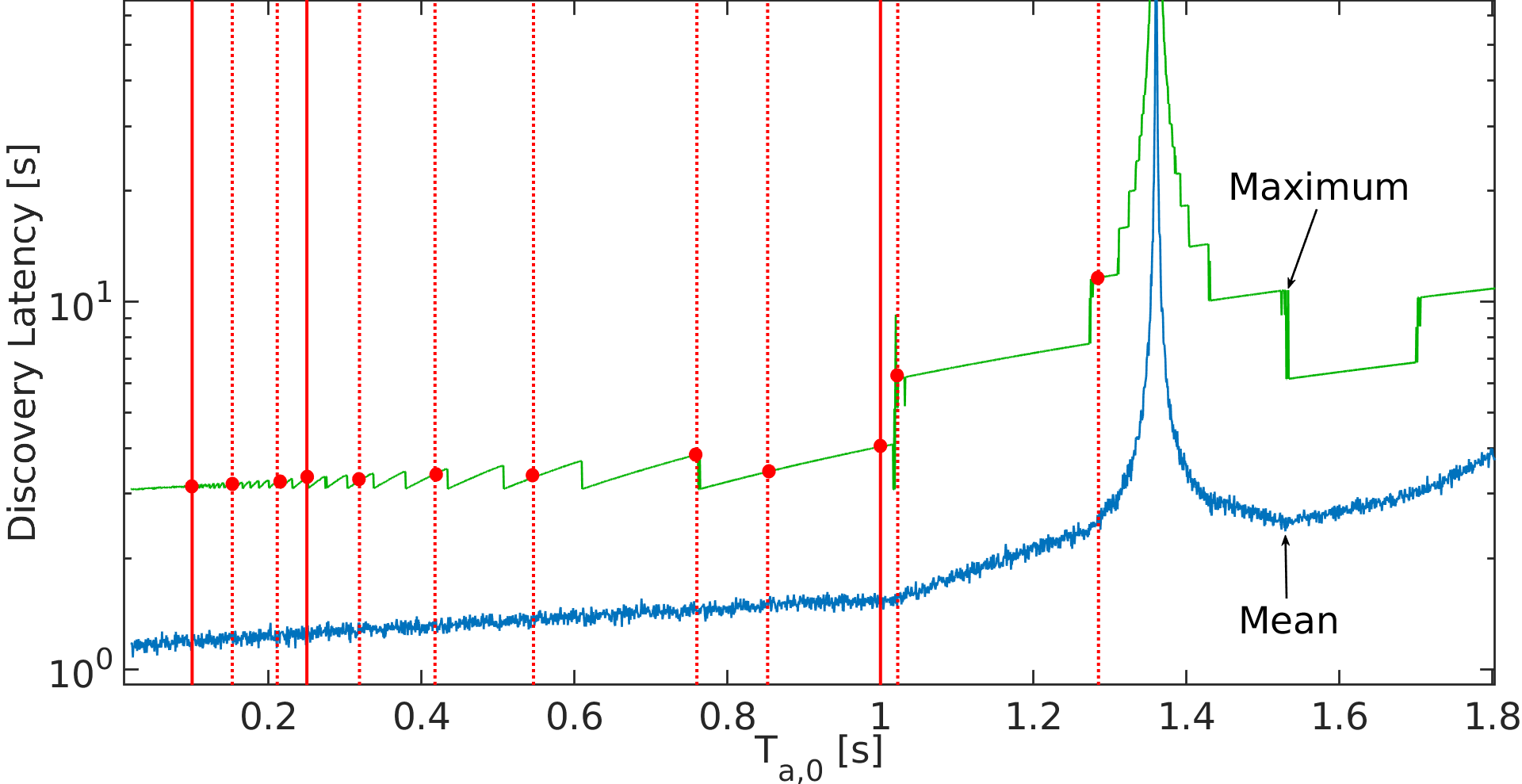}
		\caption{SCAN\_MODE\_BALANCED}
        \label{fig:latenciesAndroidBalanced}		
	\end{subfigure}
    \caption{Simulation of maximum and mean discovery latencies in Android for different values of $T_{a,0}$. The vertical lines depict values of $T_{a,0}$ used by Android (solid lines) and iOS (dashed lines).}
	\label{fig:latenciesAndroid} 
\end{figure*}

For the simplicity of exposition, in the following, we consider the discovery latency measured from the first beacon sent after two devices have been brought within range. Since the first beacon might have been sent by up to $T_{a}$ time-units after this, the worst-case latency observed in practice is by $T_{a}$ time-units larger.
Figure~\ref{fig:latenciesAndroidLowPower} depicts the simulated discovery latencies when the 
SCAN\_MODE\_LOW\_POWER setting is used, which translates to $T_s = \SI{5120}{ms}$ and $d_s = \SI{512}{ms}$. Similarly, Figure~\ref{fig:latenciesAndroidBalanced} depicts the discovery latency for the SCAN\_MODE\_BALENCED setting ($T_s = \SI{4096}{ms}$, $d_s = \SI{1024}{ms}$). We have considered all values of $T_{a,0}$ supported by \ac{BLE} in the depicted range. For every value of $T_{a,0}$, we have carried out $1,000$ simulations. For each of them, we have computed the maximum latency and the mean latency. The solid red vertical lines depict the values of $T_{a,0}$ supported by Android, and the dashed lines correspond to those supported by iOS.

Let us first consider the SCAN\_MODE\_LOW\_POWER setting.
As can be seen, for values with $T_a < d_s$, the maximum latency is approximately $\SI{5}{s}$.
However, for $T_a > d_s$, some parameterizations lead to high maximum and mean latencies. For example, for $T_{a,0} = \SI{1022.5}{ms}$, which is supported by Android, the maximum resulting latency is $\SI{172.5}{s}$. For the purpose of contact tracing, such a latency could be unacceptable, since a close contact of less than \SI{3}{minutes} could already be relevant for a virus transmission.
Recall from our previous discussion that because of issues such as resource sharing, smartphones might also deviate from the given parameter values. For example, if, due to scheduling conflicts, the value of $T_{a,0} = \SI{1022.5}{ms}$ would be slightly reduced to $\SI{1018.8}{ms}$, the discovery latency would converge towards infinity (cf. Figure~\ref{fig:latenciesAndroidLowPower}), and the devices would therefore never discover each other in most cases. Also for the parameterizations with $T_a < \SI{512}{ms}$, there are no guarantees that the depicted latency is always reached. For example, for $T_{a,0} = \SI{417.5}{ms}$, when assuming that every second beacon was dropped, the effective interval would be increased to $\SI{835}{ms}$, leading to latencies of around $L =$~$\SI{30}{s}$. For the SCAN\_MODE\_BALANCED setting, the overall situation looks similar (cf. Figure~\ref{fig:latenciesAndroidBalanced}), but a larger fraction of the values for $T_{a,0}$ lead to latencies blow $\SI{5}{s}$.
In summary, many of the supported configurations provide practical latencies of up to $\sim\SI{5}{s}$, while there are also combinations of parameters that might lead to latencies being unacceptable for contact tracing. Therefore, the settings should be chosen carefully.

In addition, as already explained, the parameter values that {\em actually} get used can not be controlled entirely by the application, especially when different tasks utilize the radio in parallel, which is the case in smartphones. Hence, in some cases, the actual latencies might differ from the ones presented above. As a result, we have to consider a smartphone as a device that maintains a certain maximum discovery latency in \textit{most} cases, while no guarantees can be given for the worst case. Finally, different smartphone manufacturers might have altered these values in their adopted versions of Android, or might do this in the future. Therefore, compatibility among all different smartphone models -- for the purpose of estimating a worst-case discovery latency -- is also not guaranteed.

\subsubsection{Energy Consumption}
\begin{table}[bth]
    \centering
    \begin{tabular}{l|l|l}
         \textbf{Scan Mode} & \textbf{nRF52832} & \textbf{BLE112}\\
         \hline
         LOW\_POWER & $0.52 - 0.57\quad\%$& $2.13 - 2.34\quad\%$  \\
         BALANCED & $1.30 - 1.35\quad\%$& $5.30 - 5.51\quad\%$ \\
         LOW\_LATENCY & $5.20 - 5.25\quad\%$& $21.14 - 21.35\quad\%$ \\
    \end{tabular}
    \caption{Reduction of the battery runtime by continuous contact tracing using different Android scan modes. Each given range of percentages accounts for all values of $T_{a,0}$ supported by Android and iOS.}
    \label{tab:batteryLifeTimes}
\end{table}
We now study how the settings from Table~\ref{tab:sttings_android} impact the energy consumption in the case of Android phones. 
We have computed the mean current draw $I_{BLE}$ of a \ac{BLE} radio that scans and advertises according to the different settings.

Different smartphones use different radios with different firmware and, accordingly, have different energy consumption. We computed the energy consumption for two \acp{SoC} with published energy data. But we did not see any smartphone using these chips. Smartphones typically use dual-mode chips that support WiFi and Bluetooth, for which, to the best of our knowledge, no data are available. For all recent devices we have investigated, these data appear to be confidential. 
Therefore, we study two different, well-chosen Bluetooth radios, and we assume that the energy consumption of the \acp{SoC} used on smartphones lie in between the energy consumption of our two chosen \acp{SoC}. First, we consider a Bluegiga BLE112 wireless module. This radio is based on the Texas Instruments CC2540 \ac{SoC}, which was among the first \ac{BLE} radios available. In addition, we consider the more recent Nordic nRF52832 \ac{SoC}.

The estimation for the BLE112 device has been carried out based on the energy model proposed in~\cite{kindt:20}. We  have thereby followed the recommendation in \cite{kindt:20} to model an advertising event (i.e., a sequence of packets on different channels used for advertising) as an event of a \ac{BLE} master in the connected mode. We have assumed that the advertiser does not listen for responses to its advertising packets, thereby simpifying the model from~\cite{kindt:20}. We have further assumed that two consecutive packets within one event on the same channel are separated from each other by \SI{150}{\micro s}. The impact of these assumptions on the energy consumption is negligible.

For the nRF52832 SoC, we have applied the energy model provided by the device manufacturer for advertising in ~\cite{nordicPowerProfiler}. For scanning, no dedicated model is available. We therefore have assumed the reception current given in its datasheet~\cite{nrf52} for estimating the energy consumption during a scan window of length $d_s$. For accounting for the switching overheads (i.e., the energy needed for  activating and deactivating the radio before and after a scan window), we have artificially extended $d_s$ by $2\times \SI{140}{\micro s}$, which is equal to 2 ramp-up times of the radio.
For both \acp{SoC}, we have assumed an operating Voltage of $\SI{3}{V}$ and a transmit power of $\SI{0}{dBm}$. Further, we have neglected the sleep current, which is small against the current draw during communication. From this data, we have computed the average current draw $I_{BLE}$ of the \ac{BLE} \ac{SoC} during contact tracing.

We have put $I_{BLE}$ in relation to the mean current draw of the smartphone. For this purpose, we have assumed that a smartphone is equipped with a battery capacity of $\SI{3000}{mAh}$. We have further assumed that this capacity is drained within $\SI{24}{h}$ when contact tracing is not carried out, which leads to an average current consumption of $I_{P} = \nicefrac{\SI{3000}{mAh}}{\SI{24}{h}}$ of the smartphone. We can now form the fraction $\nicefrac{I_{BLE}}{I_{P}}$, which gives us the percentage of time by which continuous contact tracing will drain the battery earlier.

Table~\ref{tab:batteryLifeTimes} depicts the results of this computation. The given range of values for each scan setting comprises all different advertising intervals $T_{a,0}$ supported by Android and iOS. The impact of the advertising interval $T_{a,0}$ is negligible, whereas the scan setting (which relates to $T_s$ and $d_s$) has a huge impact on the energy consumption. For the BLE112 radio, the SCAN\_MODE\_LOW\_LATENCY setting reduces the battery lifetime by about $\SI{20}{\percent}$, which will be noticable in practice. The SCAN\_MODE\_BALANCED setting reduces the battery lifetime by about $\SI{5}{\percent}$, and the SCAN\_MODE\_LOW\_POWER by about $\SI{3}{\percent}$. For the nRF51822 radio, the energy consumption is by approximately $\SI{75}{\percent}$ lower in all modes of operation. 
We can conclude that the SCAN\_MODE\_BALANCED and the SCAN\_MODE\_LOW\_POWER settings come with a sufficiently low energy-overhead, while the SCAN\_MODE\_LOW\_LATENCY can potentially reduce the battery runtime by a noticeable degree on some smartphone models.

It needs to be mentioned that compared to a \ac{ND} solution with optimal energy efficiency (cf.~\cite{kindt:19}), the parameterizations supported by Android and iOS require significantly more energy for providing the same worst-case latency. In particular, the partitioning of the overall duty-cycle $\beta + \gamma$ into reception and transmission is far from the optimum, since here, $\gamma >> \beta$, whereas $\beta \approx \gamma$ leads to the best trade-off between energy and latency. Clearly, $d_s$ could be significantly reduced for many of the supported choices of $T_{a,0}$, thereby guaranteeing the same latency with lower energy consumption. However, the overall energy demand of the Bluetooth radio is small compared to the capacity of batteries in most currently available smartphones.  As a result, the energy-demanding configurations are of no significant consequence for the battery runtime. In addition, many of these parameterizations lead to multiple beacon transmissions per scan window, which improves the resilience against collisions. We next study this issue of resilience. On the other hand, a dedicated device for contact tracing could achieve significantly higher battery lifetimes of up to 6 months, as we describe in Section~\ref{sec:wearable}.

\subsubsection{Collision Behavior}
\label{sec:collisions}
If two or more devices transmit a beacon at the same time, these beacons will overlap and hence collide. With high probability, these beacons will then not be received successfully, even when they coincide with a scan window. This might increase the discovery latency, or even prevent discovery in some cases. Note that the 3-channel-operation of \ac{BLE} does not mitigate this problem, since if a pair of beacons from two devices overlaps on the same channel, the remaining two beacon pairs on the other channels will also overlap.

The choice of $T_{a,0}$ also impacts the collision rate, since it affects the fraction of time $\beta$ during which the channel is busy. Recall that in a pessimistic contact tracing scenario, up to 75 people can come into spatial vicinity. To cover the absolute worst case, we consider that the range of the radio will exceed the one actually required for contact tracing. Hence, we assume that up to 100 devices can interfere with each other. 
\begin{figure}[b]
	\centering
	\includegraphics[width=\linewidth]{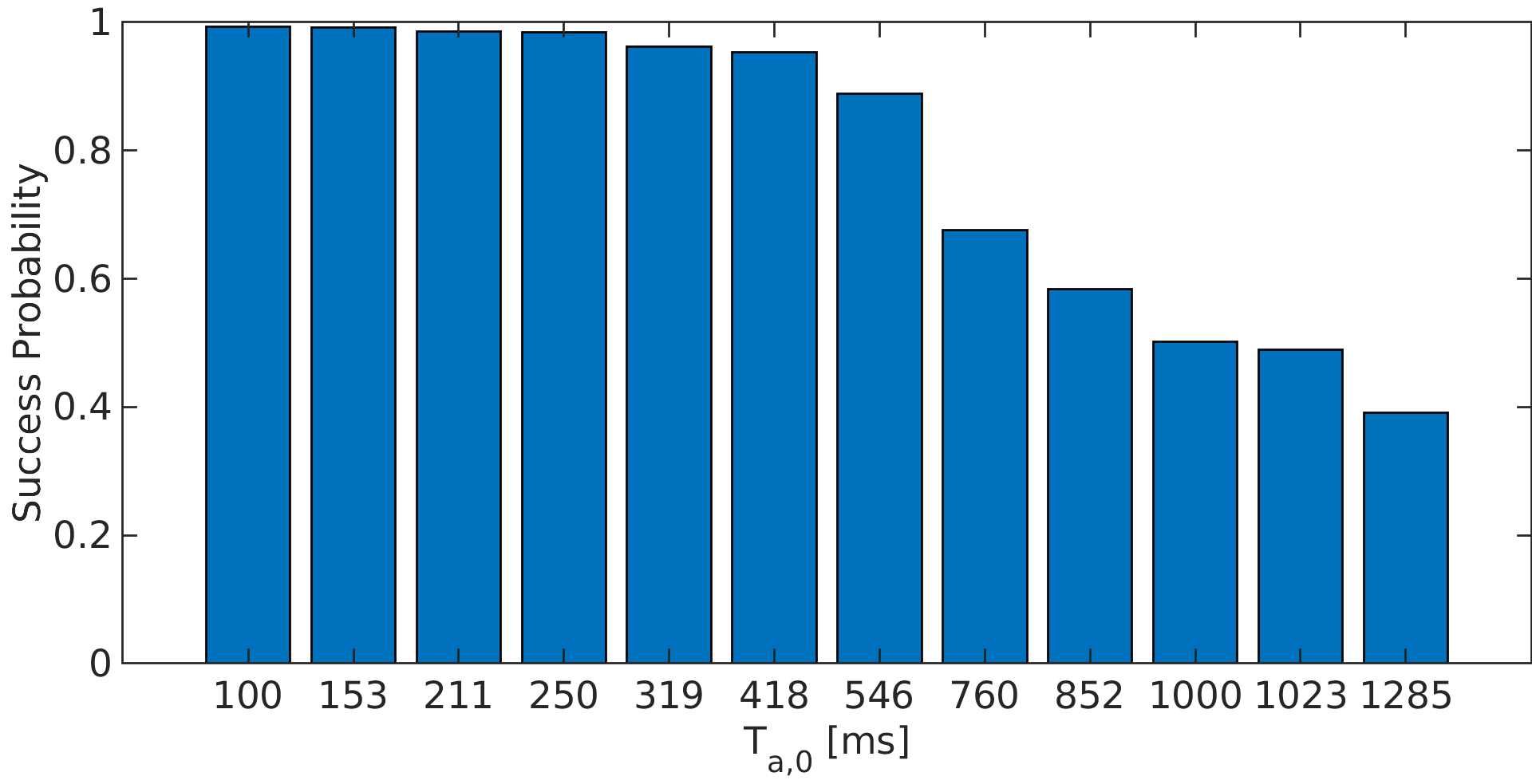}
	\caption{Probabiltiy of discovery in a crowd of 100 subjects for the SCAN\_MODE\_LOW\_POWER setting.}
	\label{fig:successprobabilities_lowPower} 
\end{figure}
We have simulated $10,000$ discovery procedures for each parameterization. We have considered a discovery procedure as successful, if at least one non-colliding beacon overlapped with a scan window within $\SI{10}{s}$. For the SCAN\_MODE\_LOW\_LATENCY and SCAN\_MODE\_BALANCED settings, assuming that every smartphone uses the same settings, the resulting success probability was $\SI{100}{\percent}$ for all values of $T_{a,0}$ supported by Android and iOS, barring one exception for SCAN\_MODE\_BALANCED with $T_{a,0} = \SI{1.285}{s}$, where the success probability was reduced to $\SI{96.4}{\percent}$. Figure~\ref{fig:successprobabilities_lowPower} depicts the probabilities for the SCAN\_MODE\_LOW\_POWER setting. As can be seen, the probabilities vary between $\SI{99.4}{\percent}$ and $\SI{39.1}{\percent}$. Hence, for some values of $T_{a,0}$, almost all discovery attempts are successful within $\SI{10}{s}$, whereas for others, more than $\sim \SI{60}{\percent}$ exceed $\SI{10}{s}$. For values of $T_{a,0} > \SI{512}{ms}$, the low success probability is caused both by colliding beacons and by the worst-case latency exceeding $\SI{10}{ms}$. 

The reasons for the high success probability for most parameterizations are a) the relatively low channel utilization $\beta = \nicefrac{\omega}{T_{a,0}}$ of each radio, and b) the large scan window $d_s$ in combination with the random delay. Since multiple beacons are sent within every scan window, the probability that one of them is received without overlapping with a beacon from a different device is high. 

\subsubsection{Summary}
\label{sec:performance_summary}
In summary, the SCAN\_MODE\_BALANCED setting results in discovery within reasonable latencies. It is compatible with almost all advertising intervals supported by Android and iOS, does not drain the battery overly quickly, and also works in crowded scenarios. The SCAN\_MODE\_LOW\_POWER setting can potentially lead to impractically high latencies and failure rates, while the SCAN\_MODE\_LOW\_LATENCY setting comes with an unnecessarily high energy consumption. For the advertising interval $T_{a,0}$, we recommend to use the ADVERTISE\_MODE\_LOW\_LATENCY or ADVERTISE\_MODE\_BALANCED settings, since this choice does not have a major impact on the energy consumption, but leads to low latencies.

From a neighbor discovery and energy (or battery life) perspective, contact tracing using smartphones appears to be feasible. However, it needs to be highlighted that there is no guarantee that a performance or reliability is always reached, and there is always a risk that certain contacts remain unrecorded. In addition, different phones and apps, e.g., iOS and Android devices, might use different parameter values, such that some smartphones can discover each other faster and with higher resilience against collisions than others. In the worst case, device manufacturers might have altered the values currently found in Android, resulting in certain phones might not being able to discover each other in a reliable manner. In other words, smartphone-based contact tracing is a {\em best-effort} mechanism without any guarantees. This should be realized by both individuals using a contact tracing app, and should also be accounted for by infection spread models that estimate the minimum adoption rates of contact tracing apps in order to contain a spread. 

As an attempt to facilitate and improve contact tracing on smartphones, Google and Apple have recently drafted a new specification~\cite{googleApple:20} for a Bluetooth contact tracing service. This draft also proposes values for $T_{a,0}$. Even when being deployed in large scale, the lack of guaranteed parameter values and hence latencies will prevail. In particular, the new specification~\cite{googleApple:20} clearly states that the advertising interval might be ``subject to change'', while being ``reccomended'' to lie between $\SI{200}{ms}$ and $\SI{270}{ms}$. Further, this specification gives no values for $T_s$ and $d_s$, but states that they need to be chosen such that a nearby device is detected within 5 minutes. Besides many contacts below 5 minutes might already be relevant (e.g., consider two persons jogging in a park or shaking hands when they meet and then disperse again), this also limits the granularity of the contact duration estimation to up to 5 minutes. As a result, such smartphone-based services need to be considered as a very coarse tool for contact tracing.

So far, we have only discussed aspects related to the neighbor discovery procedure on smartphones. In the next section, we will study additional aspects and then refine our conclusion.

\subsection{Other Limitations}
\label{sec:limitations}
Even though contact tracing on smartphones appears to be feasible from a neighbor discovery perspective, there are nevertheless multiple additional limitations that significantly reduce the performance and reliability of smartphone-based contact tracing. In particular, the following problems are inherent to smartphone-based solutions.
\begin{asparaenum}
\item For contact tracing, it is of fundamental importance to estimate the distance or proximity between two subjects. This can either be done by limiting the transmit power, such that only relevant distances are covered, or by evaluating the \ac{RSSI}.
In both cases, the distance estimation is based on the fact that wireless signals are subjected to a distance-dependent attenuation, and on the assumption that the distance can be estimated from the measured attenuation. In this estimation, there are many sources of errors. For example, the attenuation depends on the smartphone model, on the orientation of the sending and receiving antennas, the environment (e.g., metal parts in vicinity) and on the wireless channel used. While some of these uncertainties can be mitigated through various techniques, the following problem will always remain. 

Whenever the wireless signal has to pass through the human body, the attenuation is much larger than in free space. For example, \cite{alomainy:07} reports an attenuation of $\SI{19.2}{dB}$ between the chest and the back of a human body.
The exact attenuation is difficult to to predict, since it depends on the tissue, frequency, individual subject and the positions of both smartphones on the human body. However, we in the following roughly estimate what the difference in attenuation in free space and in the human body implies for distance estimation.

In~\cite{chowdhury:16}, the attenuation between a sender and a receiver placed on different parts of the human body has been measured experimentally. In some experiments, both devices were in line of sight, whereas in others, the body was in between them. Based on the interpolated RSSI values from these experiments, we exemplify the distances in free space and in tissue that lead to the same attenuation in Table~\ref{tab:distances}. For example, when the signal has to travel through $\SI{32}{cm}$ of tissue, a distance of way more than $\SI{1}{m}$ would be estimated, since the underlying model assumes that the attenuation was caused by a certain distance in free space. Hence, in many situations, the smartphone would classify a contact person as ``far away'', while actually being close. For example, consider two persons walking side-by side and holding their hands, but wearing their smartphones in opposite pockets. Though this would represent a relevant contact, the obstructed line of sight would lead to a large estimated distance, an hence to a miss-classification of the contact.
We could further confirm this behavior using our own experiments: When placing two identical smartphones on a table in a distance of around $\SI{10}{cm}$, when running the ITO demonstration app for contact tracing~\cite{ito:20}, the estimated distance was around $\SI{50}{cm}$ when both smartphones were within line of sight. When the arm of a human was placed between them, the estimated distance increased to almost $\SI{5}{m}$. 
\begin{table}
\centering
\begin{tabular}{c|c}
	\textbf{Distance Tissue} & \textbf{Distance Free Space}\\
	\hline\\
	$\SI{20}{cm}$ & $\SI{60}{cm}$ \\
	$\SI{25}{cm}$ & $\SI{1}{m}$\\
	$\SI{32}{cm}$ & $>> \SI{1}{m}$\footnote{No value for free space was given in~\cite{chowdhury:16} for the corresponding tissue attenuation.}
\end{tabular}
\caption{Distances in human tissue and the corresponding estimated free-space distance.}
\label{tab:distances}
\end{table}
\item A smartphone might not be carried on the body. For example, it might be placed in a handbag, or might be left e.g., in a car, where contacts of the device do not correspond to actual contacts of the person. This prevents a reliable reconstruction of contacts, since it will lead to missed contacts and false positives.
\item For tracing all relevant contacts using a wireless solution, it is important that as many people as possible participate. However, not everyone owns a smartphone. For example, only $\SI{81}{\percent}$ of all Germans had a smartphone in 2018~\cite{ametsreiter:19}. Furthermore, a certain part of the population may be physically or mentally unable to handle a smartphone, e.g., small children or elderly. Hence, a certain fraction of people will not be able to participate in smartphone-based contract tracing. As a result, a certain fraction of relevant contacts remain unnoticed.
\item As mentioned earlier, for contact tracing on smartphones, an application (app) needs to be installed, Bluetooth permissions need to be granted and the app needs to be activated. This will always cause handling errors, such that the tracing application is not always properly activated when people are in public. 
\item As we have already shown, the \ac{ND} procedure on smartphones consumes more energy than necessary. Though the reduction of the battery lifetime is less than about $\sim5 - 25 \%$, smartphones need to be recharged more frequently and the total energy demand of all smartphones increases.
\item Though privacy-conserving approaches are being proposed, the high amount of personal data and the various built-in sensors (e.g., microphones, cameras and GPS) will always cause privacy-concerns of smartphone users. If the application exploits its granted permissions, it might forward sensitive data to a server operated by public authorities. 
\item We have shown that the tuples of $(T_{a,0}, T_s, d_s)$ that are currently supported by recent Android and IoS smartphones are essentially compatible with each other. However, operating system developers could change them at any time. In fact, \cite{siva:19} reports a different value for the scan window $d_s$ in the  SCAN\_MODE\_BALANCED setting than what is found in the most recent Android source code. This indicates that it has been changed recently. As a result, different devices could become incompatible with each other in the future. 
\end{asparaenum}
Unfortunately, these issues are fundamental and cannot be resolved using smartphones. We therefore propose a wearable solution for contact tracing in the next section.
\section{Contact Tracing using Wearables}
\label{sec:wearable}
In this section, we for the first time propose a wearable solution and outline how it could mitigate the problems described in the previous section.
\begin{figure}[tb]
	\centering
	\includegraphics[width=\linewidth]{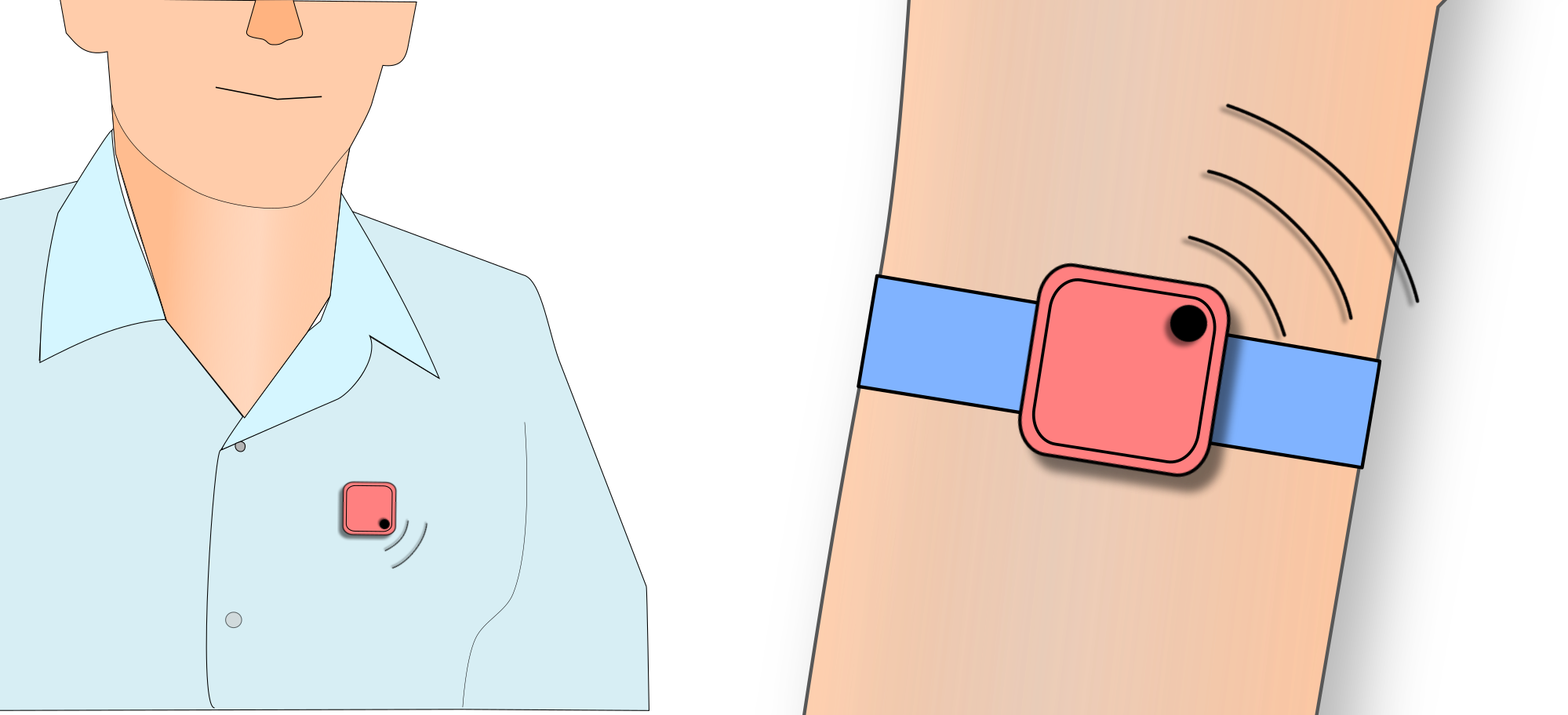}
	\caption{Wearable for contact tracing.}
	\label{fig:bracelet} 
\end{figure}
Consider for example a wrist-worn bracelet, as depicted in Figure~\ref{fig:bracelet}. Alternatively, it could also be worn as a necklace or sticker, as also shown in the figure. It would consist of simple and low-cost hardware, e.g., a wireless radio, an accelerometer and a battery. For example, let us assume the following scenario, in which such a wearable would be helpful.

In the onset of a pandemic, every person (e.g., in a hospital, an airplane, a municipality in quarantine or an entire country) obtains a wearable. While being handed out, it is registered to the corresponding person. This can be done efficiently e.g., by swiping a NFC-based passport over the device, or by simply linking its serial number and owner in a database. The wearable is always worn in public and detects relevant contacts. For assessing the relevance of a contact, a score is computed based on e.g., the contact duration, proximity and detected activity. Once a person is infected, their device's memory is read out and all IDs of devices that have been in contact throughout the incubation period are identified. The corresponding persons are then looked up from an external database. In this scenario, the mapping of IDs and persons needs to be known to a central authority. In an alternate solution, the wearables are opportunistically connected to the Internet through smartphones, using which a list of detected contacts is uploaded to a server whenever the smartphone is nearby. Here, the IDs do not need to correspond directly to persons, since in case of an infection, notifications could be sent to the smartphones of contact persons without identifying the smartphone owners. Such a wearable could be designed for economic manufacturability. Alternatively, an existing wearable used for fitness tracking could be re-used, since the contact tracing functionality can be realized in software. 

A wearable solution would mitigate or solve most of the problems associated with smartphone-based solutions described in the previous section. In particular, its advantages are as follows. \\

\par\noindent\textbf{Distance Estimation: } A bracelet always faces either the front of the body or its side. Situations in which the arm is in front of the body can be detected reliably using accelerometer data and occur regularly~\cite{kindt:15b}. Similarly, a necklace or sticker is always worn in front of the body. This greatly improves the reliability of contact detection, especially when detecting face-to-face contacts, in which droplet infections are more likely. In addition, techniques such as time-of-flight~\cite{Lanzisera:11} can be used. Here, the time within which a signal has traveled between sender and receiver is used for distance estimation instead of the attenuation. Though the latest version of the Bluetooth standard also supports this, there are no compatible smartphones, yet. Therefore, it is not clear when this will become available on the majority of smartphones. Finally, other frequency bands with more beneficial penetration properties can be used by a wearable. Also collaborative approaches, where multiple devices jointly estimate their distances for achieving a higher accuracy, are feasible. With all of these techniques, the accuracy of distance estimation could improve considerably, which could lead to a more reliable tracing.\\

\par\noindent\textbf{Adjustable Transmit Power: } While smartphones only provide a few predefined settings, the software on a wearable has full control over the radio and hence its transmit power. This means that the range of transmission can be adjusted to the range needed in the current situation in a fine-grained manner. This reduces the likelihood of collisions and further simplifies the distance estimation.\\

\par\noindent\textbf{Reliability of Operation: }
A wearable does not need any handling, since it is operational whenever it is worn. Hence, there is no possibility of faulty handling. Since it is designed to be always worn, the risk that it is left e.g., in a car, is reduced. Furthermore, since all parameters of the ND procedure can be optimized, the operation will not be as energy-expensive as on a smartphone.
For example, consider a low-cost nRF52832 radio. According to~\cite{nordicPowerProfiler}, it consumes $\SI{6.7}{mA}$ for transmitting with a power of $\SI{0}{dBM}$ and $\SI{6.7}{mA}$ also for receiving or idle-listening. We conservatively assume that every wake-up comes with an overhead of $\SI{280}{\micro s}$, during which the full reception or transmission current is consumed. This overhead corresponds to $2 \times$ the time the radio needs to switch between reception and transmission, and is hence very conservative. 

We further assume that the wearable is equipped with a $\SI{200}{mAh}$ battery, which is often found in smartwatches. For guaranteeing a discovery latency of $\SI{5}{s}$, using the parameterizations from \cite{kindt:20b}, a battery runtime of $\SI{76}{days}$ can be achieved. Though some additional measures against collisions would slightly reduce this runtime, other measures could drastically extend it. For example, a light-weight activity detection could switch off the device during sleep or when not being worn. This could extend the runtime to up to 6 months, thereby making any recharging of the battery unnecessary throughout an entire virus outbreak. As a result, the device would always be operational, which leads to an improved prevalence of operational tracing devices in the population.
Furthermore, the tuple $(T_s, T_a, d_s)$ would always be compatible with the wearables of all other persons, since all devices would have the same firmware. Finally, unlike on smartphones, the parameters actually used would not vary over time, thereby further improving the tracing reliability.\\

\par\noindent\textbf{Privacy: }
The wearable only contains the hardware needed for tracing, i.e., a radio and an accelerometer for activity detection. In addition, it would not store any private data. Further, it does not maintain a permanent Internet connection. Hence, it does not record and dispatch any unnecessary data. This greatly improves the privacy and trustworthiness compared to a smartphone. \\

\par\noindent\textbf{Detection of Physical Contacts: }
While the distance between two people is relevant for droplet infections, it could also be important to detect physical contacts (e.g., two persons shaking hands) for smear infections. It has been shown previously that electromagnetic signals can be used to modulate information over human tissue, e.g., in ~\cite{hachisuka:03}. Such a technique could also be used for detecting whether two persons have touched each other. For
example, a bracelet could both emit and try to detect such signals on the skin. Once a signal from another device can be detected, there has been a physical contact.\\
\par\noindent
In summary, a wearable device would solve many of the problems inherent to smartphone-based tracing solutions and lead to a much higher tracing reliability. 
If every person would wear such a wearable in public, it seems feasible that they would provide the evidence to inform on when lockdowns are necessary or can be dispensed with. We therefore believe that the development of such wearables is an important component in the toolbox that needs to be developed as a precaution against the spread of diseases.
\balance

\section{Concluding Remarks}
\label{sec:concluding_remarks}
In this paper, we have studied the technical feasibility of using smartphones for contact tracing. Even though their Bluetooth radios support the essential features necessary for contact tracing, tracing reliability will always be limited, potentially leading to false positives and/or missed contacts. Indeed it has been argued that to be effective, delays in notifying contacts and entry into quarantine should be minimized. Also, to prevent transmission, robust and reliable contact tracing apps or devices would need to be used by at least $\SI{60}{\percent}$ of the population \cite{ferretti:20}, even when the reliability of contact detection is $\SI{100}{\percent}$.
Hence, to realize the full potential of wireless contact tracing, approaches/devices other than smartphones need to be pursued.
As we suggest here, a wearable device can potentially overcome many of the limitations inherent to smartphones, thereby allowing for more fine-grained and reliable tracing, while at the same time reducing many of the privacy concerns that are currently being discussed. 
Through unambiguous identification of only those at risk, the necessity for mass quarantines would reduce, increasing public benefits and improving the quality of life during epidemics.

\begin{acronym}
	\acro{BLE}[BLE]{Bluetooth Low Energy}
    \acro{CRC}[CRC]{cyclic redundancy check}
    \acro{SoC}[SoC]{system-on-a-chip}
	\acro{ND}[ND]{neighbor discovery}
	\acro{RSSI}[RSSI]{received signal strength indicator}
\end{acronym}

\bibliographystyle{abbrv}
\bibliography{literature}

\end{document}